\documentclass[submission,copyright,creativecommons]{eptcs}
\providecommand{\event}{ACL2 2023} 

\usepackage{iftex}

\ifpdf
  \usepackage{underscore}         
  \usepackage[T1]{fontenc}        
\else
  \usepackage{breakurl}           
\fi
\usepackage{cite}
\usepackage{graphicx}
\usepackage{amsmath,amssymb,amsthm}
\usepackage{bbm}
\usepackage{paralist}
\usepackage[linesnumbered,ruled,vlined]{algorithm2e}
\usepackage[dvipsnames,table]{xcolor}
\usepackage{color,xcolor}
\usepackage{dsfont}
\usepackage[T1]{fontenc}
\usepackage[scaled]{beramono}
\usepackage{lstautogobble}
\usepackage{syntax}
\usepackage{dsfont,pifont}
\usepackage{float}
\usepackage{orcidlink}

\newcommand{\Comment}[1]{}
\usepackage{caption}
\usepackage{subcaption}
\usepackage{hyperref}

\newcommand{\hpc}{CPC}

\newcommand{\eg}{\emph{e.g.}}

\newcommand{\proofbuilder}{proof-builder}
\newcommand{\rules}{\varname{Rules}}
\newcommand{\hyps}{\varname{Hyps}}

\newcommand*{\funcfont}{\fontfamily{lmss}\selectfont}
\newcommand*{\codefont}{\ttfamily\small}

\newcommand\mapp{\ensuremath{\mathbin{+\mkern-10mu+}}}
\SetKwInput{KwInput}{Input}
\SetKwInput{KwOutput}{Output}
\SetKwFunction{ProveUsingHints}{ProveUsingHints}
\SetKwProg{Fn}{Function}{}{}
\SetKwFunction{CheckInductiveProof}{CheckInductiveProof}
\SetKwFunction{CheckNonInductiveProof}{CheckNonInductiveProof}
\SetKwFunction{CheckProofStep}{CheckProofStep}
\SetKwFunction{NonInductiveTranslate}{NonInductiveGenerate}
\SetKwFunction{InductiveTranslate}{InductiveGenerate}
\SetKwFunction{IndObsAndNames}{IndObsAndNames}
\SetKwFunction{DoProof}{DoProof}

\makeatletter
\DeclareRobustCommand{\varname}[1]{\begingroup\newmcodes@\mathit{#1}\endgroup}
\makeatother

\DeclareTextFontCommand{\funcfontify}{\funcfont}
\DeclareTextFontCommand{\codefontify}{\codefont}

\SetCommentSty{mycommfont}

\newcommand{\codify}[1]{\ensuremath{\mbox{\codefontify{#1}}}}
\newcommand{\xdoc}[2]{\cite[\href{https://www.cs.utexas.edu/users/moore/acl2/manuals/current/manual/?topic=#2}{\codify{#1}}]{xdoc}}

\lstdefinelanguage{handproofchecker-lang}
{
  morekeywords={Conjecture, Exportation, Contract, Completion, Context,
    Derived, Goal, Proof, Lemma, property, Prop, QED, definec, defthm},
  morecomment=[l]{;;},
}

\lstdefinestyle{proofStyle}{
language=handproofchecker-lang,
stringstyle=\ttfamily\scriptsize,
basicstyle=\ttfamily\scriptsize,
commentstyle=\ttfamily\itshape\color{gray},
tabsize=2,
showspaces=false,
showstringspaces=false,
backgroundcolor=\color{Cyan!5},
autogobble=true
}

\lstdefinestyle{proofStyleRegular}{
language=handproofchecker-lang,
stringstyle=\ttfamily\scriptsize,
basicstyle=\ttfamily\scriptsize,
commentstyle=\ttfamily\itshape\color{gray},
tabsize=2,
showspaces=false,
showstringspaces=false,
backgroundcolor=\color{Cyan!5},
autogobble=true
}

\graphicspath{ {./figs/} }

\title{Proving Calculational Proofs Correct}
\author{Andrew T. Walter \qquad\qquad Ankit Kumar \qquad\qquad Panagiotis Manolios
  \institute{Khoury College\\
    Northeastern University\\
  Massachusetts, USA}
\email{walter.a@northeastern.edu \quad\qquad kumar.anki@northeastern.edu \quad\qquad p.manolios@northeastern.edu}
}
\def\titlerunning{Proving Calculational Proofs Correct}
\def\authorrunning{A.T. Walter, A. Kumar, P. Manolios}
\begin{document}
\maketitle

\begin{abstract}
  Teaching proofs is a crucial component of any undergraduate-level
  program that covers formal reasoning. We have developed a
  calculational reasoning format and refined it over several years of
  teaching a freshman-level course, ``Logic and Computation'', to
  thousands of undergraduate students. In our companion
  paper~\cite{arxiv-hpc-paper}, we presented our calculational proof
  format, gave an overview of the calculational proof checker (CPC)
  tool that we developed to help users write and validate proofs,
  described some of the technical and implementation details of CPC
  and provided several publicly available proofs written using our
  format. In this paper, we dive deeper into the implementation
  details of CPC, highlighting how proof validation works, which helps
  us argue that our proof checking process is sound.
\end{abstract}

\section{Introduction}
\label{sec:intro}

Calculational Proof Checker (\hpc) is a tool designed to help teach
undergraduate computer science students how to write proofs. In a
previous work~\cite{arxiv-hpc-paper} we presented the calculational
proof format used by \hpc\ and gave an overview of its design and
implementation. Here we provide additional details about \hpc's
implementation and provide an argument for \hpc's soundness.

\hpc\ was designed for Manolios' freshman-level CS2800 ``Logic and
Computation'' course~\cite{lc}, which uses the ACL2 Sedan (ACL2s)
theorem prover~\cite{dillinger-acl2-sedan,acl2s} to introduce logic
and formal reasoning. ACL2s extends ACL2 with several additional
features, including the \codify{defdata} data definition
framework~\cite{defdata}, the \codify{cgen} counterexample generation
framework~\cite{chamarthi-integrating-testing,cgen,harsh-fmcad,harsh-dissertation},
a termination analysis system using calling context graphs~\cite{ccg}
and ordinals~\cite{ManoliosVroon03, ManoliosVroon04, MV05} and
property-based modeling and analysis. As nearly anyone who has taught
a formal reasoning class can attest to, teaching students how to
identify what is a proof and what is not is challenging, and teaching
students how to write proofs is even more so. The choice of proof
format is highly impactful from a pedagogical standpoint, and
therefore we put substantial effort into developing ours based on many
years of experience teaching CS2800. The proof format we use in CS2800
is heavily inspired by the calculational proof style popularized by
Dijkstra~\cite{diS90,ewds}. Dijkstra's proof format is appropriate
here because (1) its linear proofs are easier to check in a local
manner (2) explicit context forces students to identify which parts of
the context are used to discharge each step and (3) it is designed for
human consumption rather than for a proof assistant, making these
skills highly transferable.

\hpc\ checks proofs in three \emph{phases}. Phases 0 and 1 are
intended to find problems with the proof in a way that is aimed at
generating actionable and high-quality feedback for the user, and were
discussed in detail in the companion
paper~\cite{arxiv-hpc-paper}. Phase 2 involves translating the proof
into one or more ACL2s theorems with proof-builder \xdoc{proof-builder}{ACL2____PROOF-BUILDER} instructions
and checking these theorems inside of ACL2s. Therefore, the soundness
of \hpc\ reduces to the soundness of the proof-builder and
ACL2s.




Our contributions include: (1) a method for translating calculational
proofs into ACL2s theorems checkable by an unmodified ACL2s instance,
(2) a proof of soundness of \hpc\ and (3) several extensions and
libraries for ACL2s we developed for \hpc. The source code for \hpc\ is
available in a public repository~\cite{repo}.

\section{Proof Format}
\label{sec:format}
We illustrate our proof format with an example proof of a conjecture,
shown in Figure~\ref{fig:example-proof}. Notice that the proof
document starts with ACL2s definitions of relevant functions required
for the proof. We use ACL2s' \codify{definec} to define functions with
input and output types. We will discuss \codify{definec} in more
detail later, but for now one should read the definition
\lstinline[style=proofStyleRegular, mathescape=true]|(definec aapp (a :tl b :tl) :tl ...)|
as the definition of a function \codify{aapp} that
takes as argument two true lists \codify{a} and \codify{b} and returns
a true list. We also use ACL2s' \codify{property} form to state an
ACL2 theorem. The first argument to that form describes type
constraints on free variables used in the form; in this case, one can
read \lstinline[style=proofStyleRegular, mathescape=true]|(property assoc-append (x :tl y :tl z :tl) <body>)| as
\lstinline[style=proofStyleRegular, mathescape=true]|(defthm assoc-append (implies (and (tlp x) (tlp y) (tlp z)) <body>)|.

In our proof format, proofs and ACL2s expressions can be arbitrarily
interleaved, allowing for example a user to define an ACL2s function,
write a \hpc\ proof about that function, and then use that proof to
justify the admission of another ACL2s function. Users can also define
helper lemmas in ACL2s using standard ACL2s proof techniques and
subsequently apply those lemmas in a \hpc\ proof, as is done in this
example with the ACL2s lemma \codify{assoc-append}.

The proof starts with a named conjecture ``\codify{revt-rrev-help}''
followed by the expression to be proved. Note that it is not relevant
to the proof checker whether one uses \codify{Lemma} versus
\codify{Conjecture}, \codify{Property} or \codify{Theorem} to name a
proof. In this case, we want to prove the conjecture using induction,
so we specify that we are doing a proof by induction and provide the
function that gives rise to the induction scheme we want to use. We
then provide a number of subproofs (Induction Case 0 through 2), one
for each proof obligation that induction using the specified scheme
gives rise to. In this case, each of these is an equational reasoning
proof, though in general any number of them could be induction proofs
instead. For each equational reasoning proof, we provide an expression
to be proved. If the expression is of the form
$A \rightarrow B \rightarrow C$ then we require that the user use the
logical rule of exportation to eliminate the nesting of implications
and state $(A \wedge B) \rightarrow C$ in the \emph{exportation
  step}. This must be done recursively, \eg\ it should apply to an
expression of that form in the antecedent of an implication. If
desired, the statement may also be transformed into an equivalent one
using propositional logic rules during this step.  A contract
completion step (which will be discussed in detail later) comes next,
if applicable. The user then writes out the proof context (the
hypotheses of the contract completed statement, if it is an
implication) as a labeled list of \emph{context items}. Additional
context items can be added in the \emph{derived context}. In Induction
Case 2, derived context items are used to derive the consequent of the
induction hypothesis so that it can be used more easily in the
proof. Each derived context item has a list of justifications. If one
can derive \codify{nil} (false) in a derived context item, there is no
need to provide the rest of the sections for that proof. The proof
\emph{goal} (the consequent of the contract completed statement if it
is an implication, or the contract completed statement itself
otherwise) is then listed, followed by a sequence of \emph{proof
  steps}. Each proof step consists of two statements separated by a
relation and a set of justifications for that step.

\begin{figure}
  \begin{minipage}{.48\textwidth}
  \begin{lstlisting}[style=proofStyle, mathescape=true]
(definec aapp (a :tl b :tl) :tl
  (if (endp a)
      b
    (cons (first a) (aapp (rest a) b))))

(definec rrev (x :tl) :tl
  (if (endp x)
      nil
    (aapp (rrev (rest x)) (list (first x)))))

(definec revt (x :tl acc :tl) :tl
  (if (endp x)
      acc
    (revt (rest x) (cons (first x) acc))))

(property assoc-append (x :tl y :tl z :tl)
    (equal (aapp x (aapp y z))
           (aapp (aapp x y) z)))

Lemma revt-rrev-help:
(implies (and (tlp x)
              (tlp acc))
         (equal (revt x acc)
                (aapp (rrev x) acc)))

Proof by: Induction on (revt x acc)

Induction Case 0:
;; Elided case where (not (and (tlp x) (tlp acc)))
QED

Induction Case 1:
(implies (endp x)
  (implies (and (tlp x) (tlp acc))	  
           (equal (revt x acc)
                  (aapp (rrev x) acc))))

Exportation:
(implies (and (tlp x) (tlp acc) (endp x))
         (equal (revt x acc)
                (aapp (rrev x) acc)))

Context:
C1. (tlp x)
C2. (tlp acc)
C3. (endp x)

Derived Context:
D1. (equal x nil) { C1, C3 }

Goal: (equal (revt x acc) (aapp (rrev x) acc))

Proof:
(revt x acc)
== { D1, Def revt }
acc
== { Def aapp }
(aapp nil acc)
== { Def rrev, D1 }
(aapp (rrev x) acc)

QED

\end{lstlisting}\hfill
\end{minipage}
\begin{minipage}{.48\textwidth}
  \begin{lstlisting}[style=proofStyle, mathescape=true]
Induction Case 2:
(implies (and (not (endp x))
  (implies
    (and (tlp (cdr x))
         (tlp (cons (car x) acc)))
         (equal (revt (cdr x) (cons (car x) acc))
                (aapp (rrev (cdr x))
                      (cons (car x) acc)))))
  (implies (and (tlp x) (tlp acc))
           (equal (revt x acc)
                  (aapp (rrev x) acc))))

Exportation:
(implies
  (and (tlp x)
       (tlp acc)
       (not (endp x))
       (implies
         (and (tlp (cdr x))
              (tlp (cons (car x) acc)))
              (equal (revt (cdr x)
                           (cons (car x) acc))
                     (aapp (rrev (cdr x))
                           (cons (car x) acc)))))
  (equal (revt x acc)
         (aapp (rrev x) acc)))

Context:
C1. (tlp x)
C2. (tlp acc)
C3. (not (endp x))
C4. (implies (and (tlp (cdr x))
                  (tlp (cons (car x) acc)))
      (equal (revt (cdr x) (cons (car x) acc))
             (aapp (rrev (cdr x))
                   (cons (car x) acc))))

Derived Context:
D1. (tlp (cdr x)) { C1, C3, Def tlp }
D2. (tlp (cons (car x) acc)) { C2, C3, Def tlp }
D3. (equal (revt (cdr x) (cons (car x) acc))
        (aapp (rrev (cdr x)) (cons (car x) acc)))
    { D1, D2, C4, MP }

Goal: (equal (revt x acc) (aapp (rrev x) acc))

Proof:
(revt x acc)
== { Def revt, C3 }
(revt (cdr x) (cons (car x) acc))
== { D3 }
(aapp (rrev (cdr x)) (cons (car x) acc))
== { Def aapp, car-cdr axioms }
(aapp (rrev (cdr x)) (aapp (list (car x)) acc))
== { Lemma assoc-append ((x (rrev (cdr x)))
                (y (list (car x))) (z acc)) }
(aapp (aapp (rrev (cdr x)) (list (car x))) acc)
== { C3, Def rrev, car-cdr axioms }
(aapp (rrev x) acc)

QED

QED

\end{lstlisting}
\end{minipage}
\caption{An example proof written in our proof format. This proof file
  is available in our repo~\cite{repo} at the path
  \codify{example/ind-examples/pass/rrev.proof}.}
\label{fig:example-proof}
\end{figure}

A simplified and compacted version of our proof grammar is shown in
Figure~\ref{fig:grammar}. For brevity we do not include the complete
grammar of our proof format, but it is available in our
repository~\cite{repo}. Our companion paper~\cite{arxiv-hpc-paper} has
more examples of proofs written for \hpc, both from CS2800 assignments
and Dijkstra's EWDs~\cite{ewds}. We recommend that interested readers
review those example proofs.

\begin{figure}
  \centering
  \setlength{\grammarparsep}{4pt minus 6pt} 
  \setlength{\grammarindent}{9em} 
  \small
  \begin{grammar}
  <ProofDocument> ::= (<Proof> | <SExpression>)+
    
  <Proof> ::= <Type> $\mathcal{V}$ : $\mathcal{E}$ [Exportation: $\mathcal{E}$]
  [Contract Completion: $\mathcal{E}$] <Body> QED

  <Body> ::= <Simple> | <Inductive>

  <Simple> ::= [Context: <Ctx>] [Derived Context: <Dtx>] Goal: $\mathcal{E}$ Proof: <Seq>
  
  <Inductive> ::= Proof by: $\mathcal{E}$ [<ContractCase>] <BaseCase>+ <InductionCase>*

  <ContractCase> ::= Contract Case $\mathds{N}$: $\mathcal{E}$ <Body> QED
  
  <BaseCase> ::= Base Case $\mathds{N}$: $\mathcal{E}$ <Body> QED
  
  <InductionCase> ::= Induction Case $\mathds{N}$: $\mathcal{E}$ <Body> QED
  
  <Type> ::= Conjecture | Property | Lemma | Theorem

  <Ctx> ::=  (C$\mathds{N}$: $\mathcal{E}$)*

  <Dtx> ::= (D$\mathds{N}$: $\mathcal{E}$)*

  <Seq> ::= $\mathcal{B}$ ($\mathcal{R}$ \{<Hint> (, <Hint>)*\} $\mathcal{B}$)*

  <Hint> ::= <Type> $\mathcal{V}$ [$\mathcal{S}$] | C$\mathds{N}$ | D$\mathds{N}$ |
  Def $\mathcal{F}$ | $\mathcal{A}$ | algebra | obvious | PL | MP
\end{grammar}
  \normalsize
  \caption{EBNF grammar for our calculational proofs where, in the
    ACL2s universe, $\mathcal{V}$ is a fresh variable or natural
    number, $\mathcal{E}$ is an expression, $\mathds{N}$ is a natural
    number, $\mathcal{B}$ is a Boolean expression, $\mathcal{R}$ is a
    binary relation on Boolean expressions, $\mathcal{S}$ is an
    association list used to represent a valid substitution,
    $\mathcal{F}$ is a valid function name, $\mathcal{A}$ is an axiom,
    and $\langle \textit{SExpression} \rangle$ is an ACL2s event
    form. PL and MP stand for Propositional Logic and Modus Ponens
    hints respectively. Items in square brackets are optional.}
\label{fig:grammar}
\end{figure}

\section{System Architecture}
\label{sec:architecture}

We summarize the architecture of \hpc\ here.  We refer interested
readers to our companion paper~\cite{arxiv-hpc-paper} for a detailed
description. The architecture of \hpc\ consists of three primary
pieces --- the user interface, the Xtext~\cite{xtext} language
support, and the ACL2s backend. Xtext is a framework for building
domain-specific languages, and automatically generates a lexer and
parser from our proof format grammar. A user submits a proof document
for checking through one of the \hpc\ interfaces, which is parsed by
Xtext and turned into an Xtext document. Next, an Xtext
\emph{validator} that we developed runs on the Xtext document,
translating it into a form that is usable by the ACL2s backend and
invoking that backend. The ACL2s backend then runs through the proof
document and reports any issues back to the Xtext validator, which
sends that information back to the user interface for reporting to the
user. A major benefit of using Xtext in this way is the ability to
associate errors detected by the backend with regions of the user's
proof document. This means that, for example if a step is determined
to be incorrect, we can produce error underlining for that step to
help the user localize the error. Xtext also allows us to provide IDE
features like syntax highlighting and code folding with minimal
additional effort.

The ACL2s backend of \hpc\ is implemented using our ACL2s Systems
Programming methodology, which we described in an ACL2 Workshop paper
last year~\cite{walter-acl2-systems-programming}. That is, the backend
is implemented mainly in ``raw Lisp'' and makes queries to ACL2s using
the API described in our paper. This allows us to use programming
constructs that are not legal in logic-mode ACL2s code, like the Common
Lisp condition system.

As the proof format's grammar (see Figure~\ref{fig:grammar})
specifies, a proof document consists of a sequence of elements, where
each element is either a proof or an event form. Here, an event form
is a call to any of the ACL2s event functions, which are a superset of
the ACL2 event functions \xdoc{events}{ACL2____EVENTS}. In our examples, the most commonly used
event forms consist of \codify{property}, \codify{defdata} and
\codify{definec}. When \hpc\ is run on a proof document, it processes
each element of the proof document in sequence, evaluating the element
in ACL2s using \codify{ld} \xdoc{ld}{ACL2____LD} if the element is an event form, or
performing proof checking and generation if the element is a
proof. Operating in this way adds some complexity but also makes \hpc\
more flexible. For example, a user can define an ACL2s function, write
a proof about that function, and then use that proof to justify the
admission of another ACL2s function. This would not be possible if we
only supported documents where a set of ACL2s expressions (or a book)
ran before all of the proofs. If an event form evaluation or a proof
check fails, \hpc\ will report an error to the user but will continue
to operate on subsequent elements inside the proof document.

As previously stated, \hpc\ performs proof checking in three
phases. Phases 0 and 1 are primarily designed to find problems in a
proof document that we can provide actionable and high-quality
feedback for. In Phase 2, \hpc\ will translate the proof file into
appropriate ACL2s theorems complete with proof-builder commands,
which are then run through ACL2s to confirm that the proofs in that
file are correct. Our soundness argument is based entirely on Phase 2.

\section{Proof Checking}
\label{sec:checking}

Our prior work~\cite{arxiv-hpc-paper} describes Phases 0 and 1 in
detail. Here we will summarize Phases 0 and 1 and describe some
relevant aspects in more detail.

Phase 0 is a quick, syntactic check of the proof document performed by
Xtext. This is provided as part of the parser that Xtext generates
from our grammar. Phase 1 is performed in the ACL2s backend, and can
itself be broken up into performing three checks. The first is to
check that the initial setup of the proof---the contract completion,
exportation, context and derived context, all of which we will discuss
in more detail later---is correct. The second is to check that all of
the steps (for a non-inductive proof) or all of the subproofs (for an
inductive proof) are correct. For a non-inductive proof, the third
step is to confirm that the conjunction of the steps is sufficient to
prove the statement under consideration. For an inductive proof, the
third step is to confirm that the subproofs constitute the proof
obligations of the induction proof that the user specified.

\subsection{Guards and Contract Completion}

Students in CS2800 are taught to reason about programs. Using ACL2s'
\codify{defdata} \xdoc{defdata}{ACL2____DEFDATA} is helpful as it is a natural way to introduce
students to contract-driven development. Students write all of their
functions using ACL2s' \codify{definec}, which requires that the user
specify the type of input arguments to the function as well as the
type that the function outputs. \codify{definec} \xdoc{acl2s::definec}{ACL2S____DEFINEC} is hooked into ACL2's
guard system \xdoc{guard}{ACL2____GUARD} in such a way that a function defined using
\codify{definec} will have guards that assert that its arguments
satisfy their specified types. Recall that the
$\mathit{guard\ obligations}$ for an expression is the sequence of
conditions that must be true to satisfy the guards for every function
call inside that expression. The \emph{function contract} for a
\codify{definec} function is the statement that if all arguments to
the function in some function call satisfy the specified input types
(the input contract for the function is satisfied), that call
evaluates to a value that satisfies the specified output type. When a
\codify{definec} form is evaluated, in addition to proving termination
like \codify{defun}, ACL2s will prove that the function's contract
holds and will perform guard verification of the function's
body. Guard verification is the process of proving that the guard
obligations of an expression hold. Note that each \codify{defdata}
type has a ``type predicate'' associated with it---a function of one
argument that evaluates to true if and only if that argument is a
member of the corresponding type. The function contract theorems that
\codify{definec} submits are suffixed with \codify{-CONTRACT} and
\codify{-CONTRACT-TP}.

We require that the statement a user is trying to prove in \hpc\ has
an empty sequence of guard obligations (equivalently, the guard
obligations are all satisfied), or if this is not the case, we require
that the user perform \emph{contract completion} on the statement
before proving it. Contract completion refers to the process of adding
appropriate hypotheses to a statement to satisfy its guard
obligations. We will also refer to the resulting statement after
contract completion as the contract completion of the original
statement. Performing contract completion on a statement of course
changes the logical meaning of the statement. From a pedagogical
standpoint, forcing users to perform contract completion helps us
highlight the correspondence between the statements being proved and
the code (the executable bodies of the functions in the
statement). The way that \codify{definec} works is relevant here---the
logical definition of a function admitted using \codify{definec}
states that a call to the function with inputs that don't satisfy the
function's input contract will evaluate to an arbitrary value 
satisfying the function's specified return type. A consequence of this
is that a \codify{definec} function cannot be expanded into its
user-provided definition unless it is known that the function's input
contract is satisfied. It is important to note that this is different from simply adding guards to a \codify{defun}, as guards do not affect either the semantics of a function definition or the theorem prover \xdoc{guard-miscellany}{ACL2____GUARD-MISCELLANY}. Enforcing that statements are contract
completed eliminates the possibility of errors or counterexamples due
to guard violations (``type errors'').

Note that the order in which hypotheses
appear in a conjunction matters, as \codify{and} in ACL2 is logically
just syntactic sugar for \codify{if} statements and the type
information that a hypothesis provides might be necessary to satisfy
the guards of a subsequent hypothesis. For example, if the original
expression was
\lstinline[style=proofStyleRegular, mathescape=true]|(implies (in e l) (consp l))| and the ACL2s definition of
\lstinline[style=proofStyleRegular, mathescape=true]|in| requires that
\lstinline[style=proofStyleRegular, mathescape=true]|(tlp l)|, the correct
contract completion of the statement is \lstinline[style=proofStyleRegular, mathescape=true]|(implies (and (tlp l) (in e l)) (consp l))|
and \textbf{not}\\ \lstinline[style=proofStyleRegular, mathescape=true]|(implies (and (in e l) (tlp l)) (consp l))|.

It would be simpler to enforce that users provide contract completed
statements at the get-go, but having users perform contract completion
inside of \hpc\ has some advantages. In particular, having both the
original statement and the contract completed statement inside of
\hpc\ allows us to check that the contract completion was done
appropriately, \eg\ that only the necessary hypotheses were added to
satisfy the guard obligations. We do not guarantee \hpc's soundness
when the user provides a \emph{non-trivial contract completion} ---
one that is not syntactically equivalent to the exported statement (if
provided) or original statement (otherwise). To be clear, the only
situation in which a non-trivial contract completion must be provided
is when the original statement has a non-empty sequence of guard
obligations, that is, there is at least one function call in the
statement with an input contract that is not provably always true. If
a user provides a non-trivial contract completion, we currently
produce a warning notifying the user of the potential for unsoundness
and recommending they update the original conjecture so that it is
contract completed.

An example of a statement with a trivial contract completion is\newline\noindent
\lstinline[style=proofStyleRegular, mathescape=true]|(implies (and (tlp x) (tlp y)) (equal (app x y) (app y x)))|.
Since the ACL2s definition of \lstinline[style=proofStyleRegular, mathescape=true]|app|
only requires that its arguments are true lists, the two antecedents ensure that the arguments to \lstinline[style=proofStyleRegular, mathescape=true]|app| are true lists and both \lstinline[style=proofStyleRegular, mathescape=true]|tlp| and \lstinline[style=proofStyleRegular, mathescape=true]|equal| have no guards, the guard obligations for this statement are trivially true and thus no antecedents need to be added during contract completion.

An example of a statement with a non-trivial contract completion is \lstinline[style=proofStyleRegular, mathescape=true]|(implies (in e l) (consp l))|, given the definition of \lstinline[style=proofStyleRegular, mathescape=true]|in| requires that its second argument is a true list. Since the guard obligations for this statement (just \lstinline[style=proofStyleRegular, mathescape=true]|(tlp l)|) are not trivially true, a non-trivial contract completion is required. In this case, \lstinline[style=proofStyleRegular, mathescape=true]|(tlp l)| must be added as an antecedent before the \lstinline[style=proofStyleRegular, mathescape=true]|(in e l)| hypothesis, so the only possible correct contract completion is \lstinline[style=proofStyleRegular, mathescape=true]|(implies (and (tlp l) (in e l)) (consp l))|.






\subsection{Proof Building Blocks}
The basic building block of an equational reasoning proof is a proof
step --- a statement that two expressions satisfy some relation,
justified by one or more hints. In general, a step in an equational
reasoning proof in our format will look like (with $\alpha$ and
$\beta$ being S-expressions and $R$ being either a relation or an
alias for a relation):

\begin{lstlisting}[style=proofStyleRegular, mathescape=true]
  $\alpha$
  $R$ { $H_1$, ..., $H_n$ }
  $\beta$
\end{lstlisting}

We say that this step is correct if and only if ACL2s can prove the
statement \lstinline[style=proofStyle, mathescape=true]|($R$ $\alpha$ $\beta$)|
under an appropriate set of hypotheses and when constrained to an
appropriate theory. As we will discuss shortly, the appropriate set of
hypotheses and the appropriate theory both are influenced by the hints
$H = \{ H_1, ..., H_n \}$ that the user provided, but also by the
context of the proof that the step is contained inside.

\subsubsection*{Hints}
\hpc\ supports several types of hints for justifying reasoning
steps. These include $\codify{C}i$ and $\codify{D}i$ which refer to
context and derived context items respectively, $\codify{def}\ \mathit{foo}$
which allows one to reference the definition of a function (allowing
one to expand a function call into its body with an appropriate
substitution) and \codify{arithmetic} which allows many kinds of
arithmetic manipulations. Some hints have aliases (for example,
\codify{arith} is an alias for \codify{arithmetic}). Other hints only
exist for readability---for example, we use \codify{MP} (\codify{Modus
  Ponens}) to indicate that a step or derived context item is
justified by the conclusion of an implication after satisfying that
implication's hypotheses, but it does not affect \hpc's checking. Each
hint for a proof step gives rise to zero or more of hypotheses
($\hyps$), $\rules$ and $\varname{Lemma\ instantiations}$, used in
proving the proof step. \emph{Rules} here refer to proved theorems in
ACL2's database, which ACL2 can automatically apply. We define
functions $\mathrm{hyps}(h)$, $\mathrm{rules}(h)$ and
$\mathrm{instances}(h)$ to be the set of hypotheses, rules and lemma
instantiations (in a format amenable to ACL2) that a hint $h$ gives
rise to, respectively.

\begin{itemize}
\item $\codify{C}i$: add the expression corresponding to the $i^\textrm{th}$ context item as a hypothesis
\item $\codify{D}i$: add the expression corresponding to the $i^\textrm{th}$ derived context item as a hypothesis
\item $\codify{def }\mathit{foo}$: enable the definition rule(s) for the function \textit{foo}
\item $\codify{cons axioms}$: enable the following rules regarding
  \codify{cons}: \codify{(:rewrite car-cons)}, \\\codify{(:rewrite
    cdr-cons)}, \codify{car-cdr-elim}, \codify{cons-equal},
  \codify{default-car}, \codify{default-cdr}, \codify{cons-car-cdr}
\item $\codify{arithmetic}$: enable the set of rules added by including the \codify{arith-5} books
\item $\codify{evaluation}$: enable all rules of type \codify{:executable-counterpart}
\item $\codify{lemma }\mathit{foo}$: add a lemma instance
  \codify{:use} hint for $\mathit{foo}$ with the given instantiation
  (if provided). This effectively instantiates the given lemma and
  adds the resulting expression as a hypothesis.
\end{itemize}


\subsubsection*{Theories}
At different times during both Phases 1 and 2, it is useful to be able
to ask ACL2s to prove a statement while limiting the types of
reasoning that it can use. One of the ways we do this is by
controlling the set of rules that ACL2s has access to. We define
theories for certain sets of rules that are used inside \hpc:

\begin{itemize}
\item \codify{arith-5-theory} is the set of rules that are added by
  including the \codify{"arithmetic-5/top"} book in a vanilla ACL2
  instance.
\item \codify{min-theory} consists of ACL2's minimal theory (which
  includes only rules about basic built-in functions like \codify{if}
  and \codify{cons}) plus \codify{(:executable-counterpart
    acl2::tau-system)}, \\ \codify{(:compound-recognizer}
  \codify{booleanp-compound-recognizer)}, and \codify{(:definition
    not)}. The former of these three rules enables ACL2 to perform
  some type-based reasoning, and the latter two are often useful for
  reasoning about propositional logic.
\item \codify{arith-theory} which consists of some basic facts about
  \codify{+} and \codify{*}.
\item \codify{type-prescription-theory} which consists of any rules of
  type \codify{:type-prescription}
\item \codify{executable-theory} which consists of any rules of type
  \codify{:executable-counterpart}.
\item \codify{contract-theory} which is \codify{min-theory} plus
  \codify{type-prescription-theory} and any rules with names ending in
  \codify{"CONTRACT"} or \codify{"CONTRACT-TP"}. The latter rules
  correspond to the function contracts for any functions admitted
  using \codify{definec}.
\item \codify{min-executable-theory} which is the union of the rules
  in \codify{min} and \codify{executable}.
\end{itemize}

\subsubsection*{Type Hypotheses}
Almost any proof involving a function defined with \codify{definec}
requires that the function's input contract is satisfied. In early
versions of \hpc, we found that this resulted in users needing to
repeatedly include justifications in their steps corresponding to
hypotheses that some free variables in the proof statement satisfy
some type predicates. Given that users already must perform contract
completion on their proof statement, this felt like an unnecessary
burden. Therefore for any step or derived context item, \hpc\ will
automatically include hints that correspond to calls of type
predicates. For example, in \codify{Induction Case 2} in the proof
example in Section~\ref{sec:format}, \codify{C1. (tlp x)},
\codify{C2. (tlp acc)}, \codify{D1. (tlp (cdr x))} and
\codify{D2. (tlp (cons (car x) acc))} are all included ``for free'' as
justifications of any proof step.





\section{Soundness}
\label{sec:translation}
Once the user has provided a proof that passes Phases 0 and 1, we
would like to translate it into an ACL2s theorem. There are two
benefits this brings: (1) we can reduce the soundness of \hpc\ to that
of ACL2s and (2) it enables one to perform a proof in \hpc\ that might
be challenging to do in ACL2s and then use the resulting theorem in
ACL2s. The second benefit is not currently exposed in a convenient way
to users of \hpc, but we believe it would be easy to implement this
feature.

It is important to note that ACL2s contains extensions to ACL2 that
require trust tags \xdoc{defttag}{ACL2____DEFTTAG} and perform
potentially unsafe modifications to ACL2. Therefore, we can only
reduce the soundness of \hpc\ to the soundness of ACL2s, not further
to the soundness of ACL2.

Our soundness theorem is as follows: given a proof $P$ without a
non-trivial contract completion and whose proof statement is $\phi$,
if \hpc\ validates $P$ then $\phi$ is a valid statement in ACL2s, given
the same ACL2s world prior to the validation of $P$. The witness for
our soundness theorem is the ACL2s theorem that proves $\phi$.

This theorem is easy to prove, as \hpc\ will validate a proof only if
it was able to prove that proof's statement in ACL2s, using the
proof-builder instructions that \hpc\ generates as described
below. Note that we make no claims about completeness---\hpc\ may
reject a proof of a valid statement.

\subsection{Proof Builder}
Generating an ACL2s statement of a \hpc\ theorem is straightforward,
but we do not want to simply hand this statement off to ACL2s for an
automatic proof---ACL2s may decide to attempt to take a different proof
approach that requires a different set of lemmas, or may just fail to
find a proof. Ultimately our goal is to determine whether or not the
user's proof is correct, so we should be able to transform it and its
justifications into a theorem that ACL2s can prove. For this reason, we
use ACL2's \proofbuilder\ functionality, which allows us to command
the theorem prover's behavior at a much lower level.

The \proofbuilder\ operates in a manner similar to an interactive
proof assistant like Coq~\cite{coqart} or
Isabelle~\cite{isabelle-hol}: there is a \emph{proof state} consisting
of a stack of goals, each of which contains a set of hypotheses and a
statement to be proved, and one provides \emph{instructions} that
operate on the goal stack. These instructions range in granularity, with
coarse instructions like \codify{prove} (attempt to prove the current
goal entirely automatically with ACL2's full power) to fine
instructions like \codify{dive} (focus on a particular subexpression
in the current statement to be proved). ACL2's
documentation provides information about many of the
available \proofbuilder\ instructions \xdoc{proof-builder-commands}{ACL2____PROOF-BUILDER-COMMANDS}.  For \hpc\ we developed several
new \proofbuilder\ instructions, many of which are variants of
existing instructions that succeed where the existing instructions
would fail. For example, \codify{:retain-or-skip} \xdoc{acl2-pc::retain-or-skip}{ACL2-PC____RETAIN-OR-SKIP} is exactly like the
built-in \codify{:retain} \xdoc{acl2-pc::retain}{ACL2-PC____RETAIN} instruction, except that it will succeed
even when all of the existing hypotheses are retained (producing no
change in the \proofbuilder\ state). Many instructions have similar
behavior that is desirable when a human is interacting directly with
the \proofbuilder, but that is not when automatically generating
instructions. These new \proofbuilder\ instructions are available in
the ACL2 distribution, inside \codify{books/acl2s/utilities.lisp}. All
of the new instructions are listed below:

\begin{itemize}
\item \codify{:claim-simple}: exactly like \codify{:claim}, except
  that it does not automatically perform hypothesis promotion on the
  newly created goal.
\item \codify{:pro-or-skip}: exactly like \codify{:pro}, except that
  it will succeed even when no promotion is possible.
\item \codify{:drop-or-skip}: exactly like \codify{:drop}, except that
  it will succeed even when there are no top-level hypotheses and no
  arguments are provided.
\item \codify{:retain-or-skip}: exactly like \codify{:retain}, except
  that it will succeed even when all of the existing hypotheses are
  retained.
\item \codify{:cg-or-skip}: exactly like \codify{:cg}, except that it
  will succeed even when the specified goal to change to is the same
  as the current goal.
\item \codify{:instantiate}: instantiate a theorem as a hypothesis
  under the given substitution.
\item \codify{:split-in-theory}: exactly like \codify{:split}, except
  that a theory can be provided to use instead of
  \codify{minimal-theory}.
\item \codify{:by}: prove a goal using exactly an existing lemma under
  a given substitution.
\end{itemize}

\subsection{Instruction Generation Algorithms}
We will now describe how we generate proof-builder
instructions for steps and derived context items, equational reasoning
proofs, and inductive proofs. In the below algorithm listings, we will
use a typewriter font face \codify{like this} to denote S-expressions
that we generate. Some additional comments on notation:

\SetKwFunction{ProveUsingHints}{ProveUsingHints}
\SetKwProg{Fn}{Function}{}{}
\SetKwFunction{NonInductiveTranslate}{EquationalReasoningTranslate}
\SetKwFunction{InductiveTranslate}{InductiveTranslate}
\SetKwFunction{IndObsAndNames}{IndObsAndNames}

\begin{itemize}
\item $x \mapp y$ denotes the sequence produced by appending the
sequences $x$ and $y$. If $y$ is a set, then it is first transformed
into a sequence by enumerating the elements of $y$ in an arbitrary
order.
\item An ACL2s statement $x$ is a \emph{type predicate call} if and
only if it is a function call with one argument and the function name
is known by \codify{defdata} to be a type predicate.
\item Let $\mathrm{hid}(x)$ be an identifier used by the proof-builder
to refer to the hypothesis corresponding to the context or derived
context item $x$.
\item Let $\mathrm{rules}(x)$ be the set of rules that a hint $x$
gives rise to.
\item Let $\mathrm{instances}(x)$ be the set of lemma instantiations
(in a format amenable to ACL2) that a hint $x$ gives rise to.
\item \IndObsAndNames{stmt, indterm} calls ACL2's proof-builder to
determine what goals are created when one tries to prove
$\mathit{stmt}$ by performing an induction on $\mathit{indterm}$. The
output is a set of tuples $(\mathit{obs}, \mathit{name})$ where
$\mathit{obs}$ is an ACL2s statement expressing one of the created
goals and $\mathit{name}$ is the name that ACL2s gave to that goal.
\end{itemize}

In the below algorithms, we elide the complexity of matching up the
names that the proof-builder gives to the hypotheses with the names of
context items that the user gave in the proof.

\subsubsection*{Hints to Instructions}
The processes of generating proof-builder instructions for a step and
for a derived context item are similar, so we will describe them
together. Figure~\ref{fig:step-and-derived-ctx} shows the general form
of a step and a derived context item. Using names from
Figure~\ref{fig:step-and-derived-ctx}, we will define the
\emph{equivalent expression} of a step to be
\lstinline[style=proofStyle, mathescape=true]|($R$ $\alpha$ $\beta$)|
and the equivalent expression of a derived context item to be
$\gamma$. Algorithm~\ref{alg:translate-hints} is the corresponding
algorithm for this process. We pass the equivalent expression of the
step or derived context item into the \codify{stmt} input of the
algorithm. Let $EE$ be the equivalent expression of the step or
derived context item in question.

\begin{figure}
\begin{lstlisting}[style=proofStyleRegular, mathescape=true]
  ;; A step
  $\alpha$
  $R$ { $H_1$, ..., $H_n$ }
  $\beta$
  ;; A derived context item
  D$n$. $\gamma$ { $H_1$, ..., $H_n$ }
\end{lstlisting}
\caption{The general form of a proof step and a derived context item.}
\label{fig:step-and-derived-ctx}
\end{figure}

The instructions that we generate should do the following: use
\codify{:claim-simple} to add a hypothesis that the equivalent
expression of the step or derived context item holds, then prove that
this hypothesis holds using the justifications that the user provided.

We first generate a \codify{claim-simple} instruction with $EE$ as the
statement to cause the proof-builder to add $EE$ as a hypothesis in
the current goal. This also results in the creation of a new goal to
prove $EE$ given the current set of hypotheses (before $EE$ was
added). We pass \codify{:hints :none} to the \codify{claim-simple}
instruction so that ACL2s does not try to prove this new goal
automatically. Then, we generate a \codify{cg} instruction to change
to the newly generated goal. Next, we calculate the guard obligations
that this goal would have given the current context and generate a
\codify{claim} instruction with those obligations as its
statement. This \codify{claim} instruction will result in ACL2s trying
to prove that the statement holds automatically. Next, based on
$H_1, ..., H_n$, we determine which context and derived context items
should be available when proving $EE$. We then generate a
\codify{retain-or-skip} instruction with appropriate arguments to only
retain the appropriate context and derived context items. We then
determine based on $H_1, ..., H_n$ what ACL2 rules should be
available. We generate a \codify{in-theory} instruction with
appropriate arguments to enable and disable rules appropriately.
Finally, we generate the instruction \codify{(:finish :bash)}, which
tells ACL2s to attempt to prove the current goal while limiting its
abilities. If ACL2s is unable to prove the goal, it will raise an error
and the proof attempt will result in a failure.

\begin{algorithm}
  \Fn{\ProveUsingHints{stmt, hints, ctx}}{
  \KwIn{$\mathit{stmt}$ is the statement to prove, $\mathit{hints}$ is the set of hints the user provided, and $\mathit{ctx}$ is the set of context and derived context items it should be proved under.}
  $I \gets []$\;
  \tcc{Add $stmt$ as a hypothesis and as a new goal, do not attempt to prove it automatically, and switch to the new goal}
    $I \gets I \mapp [\codify{(:claim-simple } \mathit{stmt} \codify{... :hints :none)}, \codify{:cg}]$\;
    $\mathit{hyps} \gets \{ x\ \vert\ x \in \mathit{hints} \wedge x \text{ is a context or derived context hint } \}$\;
    $\mathit{contracts} \gets \mathrm{GO}((\bigwedge_{h \in \mathit{hyps}} h) \rightarrow \mathit{stmt})$\;
    \If{$\mathit{contracts} \neq \codify{true}$}{
      \tcc{Add $\mathit{contracts}$ as a hypothesis and as a new goal \& prove it automatically.}
      $I \gets I \mapp [\codify{(:claim } \mathit{contracts} \codify{...)}]$\;
    }
    $\mathit{typectx} \gets \{ x\ \vert\ x \in \mathit{ctx} \wedge x \text{ is a type-predicate call } \}$\;
    $\mathit{contractsidx} \gets \text{a set containing the identifier of the } \mathit{contracts} \text{ hypothesis if } \mathit{contracts} \neq \codify{true} \text{ or } \emptyset \text{ otherwise}$\;
    \tcc{Only keep the hypotheses we should have given the hints the user provided}
    $I \gets I \mapp [\codify{(:retain-or-skip } \{ \mathrm{hid}(x)\ \vert\ x \in \mathit{hyps} \cup \mathit{typectx} \cup \mathit{contractsidx} \} \codify{)}]$\;
    $\mathit{hintrules} \gets \bigcup\ \{ \mathrm{rules}(x)\ \vert\ x \in \mathit{hints} \}$\;
    \tcc{Ensure that only the rules that we should have access to given the user's hints are available}
    $I \gets I \mapp [\codify{(:in-theory (union-theories (theory 'contract-theory) } \mathit{hintrules} \codify{))}]$\;
    $\mathit{lemmainstances} \gets \bigcup\ \{ \mathrm{instances}(x)\ \vert\ x \text{ is a hint for } Dx_i \}$\;
    \tcc{Add any lemma instances that the user described in the hints}
    $I \gets I \mapp \{ \codify{(:instantiate } x \codify{)}\ \vert\ x \in \mathit{lemmainstances} \}$\;
    \tcc{Ask ACL2 to automatically prove this goal without induction, and then reset to the original theory}
    $I \gets I \mapp [\codify{(:finish :bash)}, \codify{:in-theory}]$\;

    \Return $I$
  }
  \caption{proof-builder instruction generation for a step or derived context item}
  \label{alg:translate-hints}
\end{algorithm}

\subsubsection*{Equational reasoning proofs}
Generating instructions for an equational reasoning proof is fairly
straightforward; the algorithm is shown in
Algorithm~\ref{alg:non-inductive-translate}.

We start with \codify{:pro-or-skip} to expand the proof statement's
implication into antecedents and a consequent (if it is an
implication). Then, we generate instructions to add a hypothesis for
each derived context item and prove that it holds given the provided
justifications. We do something very similar for each proof
step. Then, we \codify{:demote} to turn the goal and hypotheses into
an ACL2 implication statement before repeatedly calling
\codify{(:split-in-theory min-executable-theory)} until the goal has
been discharged. We use \codify{:split-in-theory} (and therefore
\codify{:split}) here as it is a convenient way to invoke ACL2 with
very limited reasoning ability (just simplification, preprocessing,
and whatever rules are in the given theory).

\begin{algorithm}
  \Fn{\NonInductiveTranslate{C, D, R, P, H}}{
  \KwIn{$C$ and $D$ are the sets of all non-derived context and
  derived context items for a proof respectively. $R$, $P$ and $H$ are the
  relations, step statements and hints for the proof's proof steps, indexed
  from the start of the proof. This function only operates on equational
  reasoning proofs.}
  $I \gets []$\;
  \tcc{Perform exportation and expand implication into hyps/conclusion}
  $I \gets I \mapp [\codify{:pro-or-skip}]$\;
  \tcc{Generate instructions for each derived context item}
  \ForEach{$i \in [1..m]$}{
    $\mathit{stmt} \gets \text{the proof statement associated with } Dx_i$\;
    \tcc{Do not include any later derived context items in the context used to prove $Dx_i$}
    $\mathit{ctx} \gets C \cup \{ Dx_j\ \vert\ j \in [1..i-1] \}$\;
    $I \gets I \mapp \ProveUsingHints{stmt, hints, ctx}$\;
  }
  \tcc{Generate instructions for each step}
  \ForEach{$i \in [1..n]$}{
    $\mathit{stmt} \gets $ \codify{(} $R_i P_i P_{i+1}$ \codify{)}\;
    $\mathit{ctx} \gets C \cup D$\;
    $I \gets I \mapp \ProveUsingHints{stmt, H_i, ctx}$\;
  }
  \tcc{Turn the hypotheses and goal into an implication}
  $I \gets I \mapp [\codify{:demote}]$\;
  \tcc{Repeatedly call :split-in-theory until the proof is successful or we reach a fixpoint}
  $I \gets I \mapp [\codify{(:finish (:repeat-until-done (:split-in-theory min-executable-theory)))}]$\;
  \Return $I$
  }
  \caption{proof-builder instruction generation for a non-inductive conjecture}
  \label{alg:non-inductive-translate}
\end{algorithm}

\subsubsection*{Inductive proofs}
The algorithm for generating proof-builder instructions for inductive
proofs is provided in Algorithm~\ref{alg:inductive-translate}.

Assume we have a proof by induction without a non-trivial contract
completion. This can be thought of as several separate proofs, one for
each induction proof obligation. For each of these subproofs, we will
generate a separate ACL2s proof, complete with proof-builder
instructions. Then, we generate an ACL2s proof for the top-level
induction proof with instructions to perform a proof by induction
using the induction scheme that the user specified, and generate
instructions that discharge each proof obligation using the
corresponding generated ACL2s proof. This approach requires that \hpc\
determine the order of the subgoals that ACL2s will generate when asked
to perform an induction proof with the given induction scheme so that
we can map up the subproofs that the user performed with these
subgoals, and thus in the instructions for each subgoal we can refer
to the appropriate generated ACL2s proof. We generate these subgoals by
using the ACL2 function \codify{state-stack-from-instructions}, which
allows one to get the state of the proof builder after running a
sequence of instructions, and then reuse some existing \hpc\ code to
find a bijection between these subgoals and the induction proof cases
that the user provided.

Once we have generated proof-builder instructions for an inductive
proof, we generate \codify{defthm}s with proof-builder instructions
for all of its subproofs. We then generate an \codify{encapsulate}
statement and insert all of the subproof \codify{defthm}s in the
encapsulate as \codify{local}. The inductive proof itself is not
inserted into a \codify{local} and is thus exported from the
\codify{encapsulate}. We do not want to export the subproof
\codify{defthm}s, as they are only needed to show that the top-level
inductive proof theorem holds.

\begin{algorithm}
  \Fn{\InductiveTranslate{M, stmt, indterm, PC}}{
    \KwIn{This function only operates on inductive proofs.
    $M$ is a function mapping the names of the proof cases
    of this inductive proof to corresponding ACL2
    theorems. $\mathit{stmt}$ is the proof statement for this
    inductive proof, and $\mathit{indterm}$ is the induction
    term. $\mathit{PC}$ is the set of proof cases given for this
    inductive proof, where $\mathrm{name}(PC_i)$ is the name of
    $\mathit{PC}_i$ and $\mathrm{stmt}(\mathit{PC}_i)$ is the proof
    statement for $\mathit{PC}_i$.}
    
  \tcc{Generate the proof obligations an induction on $\mathit{indterm}$ will give rise to}
  $\mathit{obsnames} \gets \IndObsAndNames{stmt, indterm}$\;
  \tcc{Perform exportation, expand implication into hyps/conclusion, perform induction}
  $I \gets [\codify{:pro-or-skip}, \codify{(:induct }\mathit{indterm}\codify{)}]$\;

  Attempt to find an injective mapping from $\mathit{obsnames}$ to
  $\mathit{PC}$, where an element $(\mathit{obs}, \mathit{name}) \in
  \mathit{obsnames}$ is mapped to an element $\mathit{PC}_i \in
  \mathit{PC}$ iff the conjunction of the hypotheses of $obs$ after
  exportation are propositionally equivalent to the conjunction of the
  hypotheses of $\mathrm{stmt}(\mathit{PC}_i)$ after exportation.\;

  \If{no such mapping exists}{
    raise an error\;
  }

  $\mathit{inj} \gets $ the injective mapping\; 

  \ForEach{$(\mathit{obs}, \mathit{name}) \in \mathit{obsnames}$}{
    $\mathit{injPC} \gets \mathit{inj}((\mathit{obs}, \mathit{name}))$\;
    \tcc{Change to the induction obligation, use the existing proof to discharge it}
    $I \gets [\codify{(:cg-or-skip }\mathit{name}\codify{)}, \codify{(:finish :demote (:by }M(\mathrm{name}(\mathit{injPC}))\codify{))}]$\;
  }
  \Return $I$
  }
  \caption{proof-builder instruction generation for an inductive conjecture}
  \label{alg:inductive-translate}
\end{algorithm}

\section{Related Work}
\label{sec:related-work}

Our previous work~\cite{arxiv-hpc-paper} contains a longer discussion
of works surrounding the use and mechanical verification of
calculational proofs. Below we provide a summary of that discussion,
as well as some related work in ACL2 in particular.

Calculational proofs were popularized by in the early 1990s by
Dijkstra and Scholten~\cite{diS90}, Gasteren \cite{vg90} and
Gries~\cite{gri91}. A series of works~\cite{rswindow, grundy96trx,
  back97, grundy96} by Robinson, Stables, Back, Grundy and Wright
resulted in the development of structured calculational proofs, an
extension of the calculational proof style that allows for the
hierarchical decomposition of proofs. This format reduces to natural
deduction, but maintains the benefits of calculational proofs while
also allowing for improved readability and browsability of proofs.

Manolios argued for the formalization of calculational proofs and
their mechanized checking in 2000~\cite{man01}. Mizar~\cite{mizar92}
is a system for checking calculational proofs first developed in the
1970s. Several systems inspired by Mizar have been developed since,
including Isabelle/Isar~\cite{isar} and Leino et. al's poC extension
to Dafny~\cite{lepo13}. These systems typically follow Mizar's format
in not requiring the user to explicitly state the proof context. Mizar
has only lightweight support for automated reasoning in proving that
proof steps hold and only allows equality relations inside of
proofs. Isar allows for arbitrary relations and provides access to
Isabelle's powerful reasoning capabilities, like \emph{simp} for Isabelle’s
simplifier, and \emph{auto} for a combination of several
tools~\cite{lepo13}. poC only allows a predefined set of relations but
is as declarative as Mizar is, while providing more powerful automated
reasoning with its SMT solver backend.



\section{Conclusion and Future work}
\label{sec:conclusion}
We have presented an argument for the soundness of \hpc, based on its
translation of calculational proofs into ACL2s theorems with
proof-builder instructions. We are interested in seeing how \hpc\ can
be used by ``professional'' users to design their proofs, and have
some ideas about functionality that would be appropriate for these
users. In particular, we see the need to provide more automation to
such users, for example automatic generation of context and induction
proof obligations or a ``bash'' mode for eliding simple subproofs like
contract cases in inductive proofs. We hope to continue to
extend and improve \hpc\ based on requests from students and the
community, and plan on continuing to use it to help teach
undergraduates how to write proofs.

\section*{Acknowledgments}
We sincerely thank Ken Baclawski, Raisa Bhuiyan, Harsh Chamarthi,
Peter Dillinger, Robert Gold, Jason Hemann, Andrew Johnson, Alex
Knauth, Michael Lin, Riccardo Pucella, Sanat Shajan, Olin Shivers,
David Sprague, Atharva Shukla, Ravi Sundaram, Stavros Tripakis, Thomas
Wahl, Josh Wallin, Kanming Xu and Michael Zappa for their help with
developing and teaching with \hpc. We also thank all of the Teaching
Assistants and students in CS2800, who are too numerous to list
individually, who used and provided feedback on \hpc.

\bibliographystyle{eptcs}
\bibliography{refs}

\begin{thebibliography}{10}
\providecommand{\bibitemdeclare}[2]{}
\providecommand{\surnamestart}{}
\providecommand{\surnameend}{}
\providecommand{\urlprefix}{Available at }
\providecommand{\url}[1]{\texttt{#1}}
\providecommand{\href}[2]{\texttt{#2}}
\providecommand{\urlalt}[2]{\href{#1}{#2}}
\providecommand{\doi}[1]{doi:\urlalt{https://doi.org/#1}{#1}}
\providecommand{\eprint}[1]{arXiv:\urlalt{https://arxiv.org/abs/#1}{#1}}
\providecommand{\bibinfo}[2]{#2}

\bibitemdeclare{misc}{xdoc}
\bibitem{xdoc}
\bibinfo{author}{\surnamestart {ACL2 Contributors}\surnameend}:
  \emph{\bibinfo{title}{ACL2 XDoc Documentation}}.
\newblock
  \bibinfo{howpublished}{\url{https://www.cs.utexas.edu/users/moore/acl2/manuals/current/manual/index.html}}.

\bibitemdeclare{article}{back97}
\bibitem{back97}
\bibinfo{author}{Ralph \surnamestart Back\surnameend}, \bibinfo{author}{Jim
  \surnamestart Grundy\surnameend} \& \bibinfo{author}{Joakim \surnamestart von
  Wright\surnameend} (\bibinfo{year}{1997}): \emph{\bibinfo{title}{Structured
  Calculational Proof}}.
\newblock {\slshape \bibinfo{journal}{Formal Aspects of Computing}}
  \bibinfo{volume}{9}(\bibinfo{number}{5-6}), pp. \bibinfo{pages}{469--483},
  \doi{10.1007/BF01211456}.

\bibitemdeclare{book}{coqart}
\bibitem{coqart}
\bibinfo{author}{Yves \surnamestart Bertot\surnameend} \&
  \bibinfo{author}{Pierre \surnamestart Cast{\'{e}}ran\surnameend}
  (\bibinfo{year}{2004}): \emph{\bibinfo{title}{Interactive Theorem Proving and
  Program Development - Coq'Art: The Calculus of Inductive Constructions}}.
\newblock \bibinfo{series}{Texts in Theoretical Computer Science. An {EATCS}
  Series}, \bibinfo{publisher}{Springer}, \doi{10.1007/978-3-662-07964-5}.

\bibitemdeclare{inproceedings}{dillinger-acl2-sedan}
\bibitem{dillinger-acl2-sedan}
\bibinfo{author}{Harsh \surnamestart Chamarthi\surnameend},
  \bibinfo{author}{Peter~C. \surnamestart Dillinger\surnameend},
  \bibinfo{author}{Panagiotis \surnamestart Manolios\surnameend} \&
  \bibinfo{author}{Daron \surnamestart Vroon\surnameend}
  (\bibinfo{year}{2011}): \emph{\bibinfo{title}{The "{ACL2}" Sedan Theorem
  Proving System}}.
\newblock In: {\slshape \bibinfo{booktitle}{Tools and Algorithms for the
  Construction and Analysis of Systems (TACAS)}},
  \doi{10.1007/978-3-642-19835-9\_27}.

\bibitemdeclare{phdthesis}{harsh-dissertation}
\bibitem{harsh-dissertation}
\bibinfo{author}{Harsh~Raju \surnamestart Chamarthi\surnameend}
  (\bibinfo{year}{2016}): \emph{\bibinfo{title}{Interactive Non-theorem
  Disproving}}.
\newblock Ph.D. thesis, \bibinfo{school}{Northeastern University},
  \doi{10.17760/D20467205}.

\bibitemdeclare{inproceedings}{chamarthi-integrating-testing}
\bibitem{chamarthi-integrating-testing}
\bibinfo{author}{Harsh~Raju \surnamestart Chamarthi\surnameend},
  \bibinfo{author}{Peter~C. \surnamestart Dillinger\surnameend},
  \bibinfo{author}{Matt \surnamestart Kaufmann\surnameend} \&
  \bibinfo{author}{Panagiotis \surnamestart Manolios\surnameend}
  (\bibinfo{year}{2011}): \emph{\bibinfo{title}{Integrating Testing and
  Interactive Theorem Proving}}.
\newblock In \bibinfo{editor}{David~S. \surnamestart Hardin\surnameend} \&
  \bibinfo{editor}{Julien \surnamestart Schmaltz\surnameend}, editors:
  {\slshape \bibinfo{booktitle}{Proceedings 10th International Workshop on the
  {ACL2} Theorem Prover and its Applications}}, {\slshape
  \bibinfo{series}{{EPTCS}}}~\bibinfo{volume}{70}, pp. \bibinfo{pages}{4--19},
  \doi{10.4204/EPTCS.70.1}.

\bibitemdeclare{inproceedings}{cgen}
\bibitem{cgen}
\bibinfo{author}{Harsh~Raju \surnamestart Chamarthi\surnameend},
  \bibinfo{author}{Peter~C. \surnamestart Dillinger\surnameend},
  \bibinfo{author}{Matt \surnamestart Kaufmann\surnameend} \&
  \bibinfo{author}{Panagiotis \surnamestart Manolios\surnameend}
  (\bibinfo{year}{2011}): \emph{\bibinfo{title}{Integrating Testing and
  Interactive Theorem Proving}}.
\newblock In: {\slshape \bibinfo{booktitle}{International Workshop on the
  {ACL2} Theorem Prover and its Applications}}, \bibinfo{series}{{EPTCS}},
  \doi{10.4204/EPTCS.70.1}.

\bibitemdeclare{inproceedings}{defdata}
\bibitem{defdata}
\bibinfo{author}{Harsh~Raju \surnamestart Chamarthi\surnameend},
  \bibinfo{author}{Peter~C. \surnamestart Dillinger\surnameend} \&
  \bibinfo{author}{Panagiotis \surnamestart Manolios\surnameend}
  (\bibinfo{year}{2014}): \emph{\bibinfo{title}{Data Definitions in the {ACL2}
  Sedan}}.
\newblock In: {\slshape \bibinfo{booktitle}{Proceedings Twelfth International
  Workshop on the {ACL2} Theorem Prover and its Applications}},
  \bibinfo{series}{{EPTCS}}, \doi{10.4204/EPTCS.152.3}.

\bibitemdeclare{inproceedings}{harsh-fmcad}
\bibitem{harsh-fmcad}
\bibinfo{author}{Harsh~Raju \surnamestart Chamarthi\surnameend} \&
  \bibinfo{author}{Panagiotis \surnamestart Manolios\surnameend}
  (\bibinfo{year}{2011}): \emph{\bibinfo{title}{Automated specification
  analysis using an interactive theorem prover}}.
\newblock In \bibinfo{editor}{Per \surnamestart Bjesse\surnameend} \&
  \bibinfo{editor}{Anna \surnamestart Slobodov{\'{a}}\surnameend}, editors:
  {\slshape \bibinfo{booktitle}{International Conference on Formal Methods in
  Computer-Aided Design, {FMCAD} '11}}, \bibinfo{publisher}{{FMCAD} Inc.}, pp.
  \bibinfo{pages}{46--53}.
\newblock \urlprefix\url{https://dl.acm.org/doi/10.5555/2157654.2157665}.

\bibitemdeclare{misc}{ewds}
\bibitem{ewds}
\bibinfo{author}{Edgar~W. \surnamestart Dijkstra\surnameend}:
  \emph{\bibinfo{title}{EWDs}}.
\newblock \bibinfo{howpublished}{\url{https://www.cs.utexas.edu/users/EWD/}}.

\bibitemdeclare{book}{diS90}
\bibitem{diS90}
\bibinfo{author}{Edsger~W. \surnamestart Dijkstra\surnameend} \&
  \bibinfo{author}{Carel~S. \surnamestart Scholten\surnameend}
  (\bibinfo{year}{1990}): \emph{\bibinfo{title}{Predicate Calculus and Program
  Semantics}}.
\newblock \bibinfo{series}{Texts and Monographs in Computer Science},
  \bibinfo{publisher}{Springer}, \doi{10.1007/978-1-4612-3228-5}.

\bibitemdeclare{inproceedings}{acl2s}
\bibitem{acl2s}
\bibinfo{author}{Peter~C. \surnamestart Dillinger\surnameend},
  \bibinfo{author}{Panagiotis \surnamestart Manolios\surnameend},
  \bibinfo{author}{Daron \surnamestart Vroon\surnameend} \&
  \bibinfo{author}{J.~Strother \surnamestart Moore\surnameend}
  (\bibinfo{year}{2007}): \emph{\bibinfo{title}{{ACL2s}: ``The {ACL2}
  Sedan''}}.
\newblock In: {\slshape \bibinfo{booktitle}{Proceedings of the 7th Workshop on
  User Interfaces for Theorem Provers (UITP 2006)}},
  \bibinfo{series}{Electronic Notes in Theoretical Computer Science},
  \doi{10.1016/j.entcs.2006.09.018}.

\bibitemdeclare{book}{vg90}
\bibitem{vg90}
\bibinfo{author}{Antonetta J.~M. \surnamestart van Gasteren\surnameend}
  (\bibinfo{year}{1990}): \emph{\bibinfo{title}{On the Shape of Mathematical
  Arguments}}.
\newblock {\slshape \bibinfo{series}{Lecture Notes in Computer Science}}
  \bibinfo{volume}{445}, \bibinfo{publisher}{Springer},
  \doi{10.1007/BFb0020908}.

\bibitemdeclare{article}{gri91}
\bibitem{gri91}
\bibinfo{author}{David \surnamestart Gries\surnameend} (\bibinfo{year}{1991}):
  \emph{\bibinfo{title}{Teaching Calculation and Discrimination: {A} More
  Effective Curriculum}}.
\newblock {\slshape \bibinfo{journal}{Communications of the {ACM}}}
  \bibinfo{volume}{34}(\bibinfo{number}{3}), pp. \bibinfo{pages}{44--55},
  \doi{10.1145/102868.102870}.

\bibitemdeclare{article}{grundy96}
\bibitem{grundy96}
\bibinfo{author}{Jim \surnamestart Grundy\surnameend} (\bibinfo{year}{1996}):
  \emph{\bibinfo{title}{A browsable format for proof presentation}}.
\newblock {\slshape \bibinfo{journal}{Logic, Mathematics, and the
  Computer--Foundations: History, Philosophy and Applications}}
  \bibinfo{volume}{14}, pp. \bibinfo{pages}{171--178}.
\newblock
  \urlprefix\url{https://www.researchgate.net/publication/2359706_A_Browsable_Format_for_Proof_Presentation}.

\bibitemdeclare{article}{grundy96trx}
\bibitem{grundy96trx}
\bibinfo{author}{Jim \surnamestart Grundy\surnameend} (\bibinfo{year}{1996}):
  \emph{\bibinfo{title}{Transformational Hierarchical Reasoning}}.
\newblock {\slshape \bibinfo{journal}{The Computer Journal}}
  \bibinfo{volume}{39}(\bibinfo{number}{4}), pp. \bibinfo{pages}{291--302},
  \doi{10.1093/comjnl/39.4.291}.

\bibitemdeclare{inproceedings}{lepo13}
\bibitem{lepo13}
\bibinfo{author}{K.~Rustan~M. \surnamestart Leino\surnameend} \&
  \bibinfo{author}{Nadia \surnamestart Polikarpova\surnameend}
  (\bibinfo{year}{2013}): \emph{\bibinfo{title}{Verified Calculations}}.
\newblock In \bibinfo{editor}{Ernie \surnamestart Cohen\surnameend} \&
  \bibinfo{editor}{Andrey \surnamestart Rybalchenko\surnameend}, editors:
  {\slshape \bibinfo{booktitle}{Verified Software: Theories, Tools, Experiments
  - 5th International Conference, {VSTTE} 2013, Menlo Park, CA, USA, May 17-19,
  2013, Revised Selected Papers}}, {\slshape \bibinfo{series}{Lecture Notes in
  Computer Science}} \bibinfo{volume}{8164}, \bibinfo{publisher}{Springer}, pp.
  \bibinfo{pages}{170--190}, \doi{10.1007/978-3-642-54108-7\_9}.

\bibitemdeclare{misc}{lc}
\bibitem{lc}
\bibinfo{author}{Panagiotis \surnamestart Manolios\surnameend}:
  \emph{\bibinfo{title}{Logic and Computation Class}}.
\newblock
  \bibinfo{howpublished}{\url{https://www.ccs.neu.edu/home/pete/courses/Logic-and-Computation/2022-Fall/}}.

\bibitemdeclare{article}{man01}
\bibitem{man01}
\bibinfo{author}{Panagiotis \surnamestart Manolios\surnameend} \&
  \bibinfo{author}{J~Strother \surnamestart Moore\surnameend}
  (\bibinfo{year}{2001}): \emph{\bibinfo{title}{On the desirability of
  mechanizing calculational proofs}}.
\newblock {\slshape \bibinfo{journal}{Information Processing Letters}}
  \bibinfo{volume}{77}(\bibinfo{number}{2-4}), pp. \bibinfo{pages}{173--179},
  \doi{10.1016/S0020-0190(00)00200-3}.

\bibitemdeclare{inproceedings}{ManoliosVroon03}
\bibitem{ManoliosVroon03}
\bibinfo{author}{Panagiotis \surnamestart Manolios\surnameend} \&
  \bibinfo{author}{Daron \surnamestart Vroon\surnameend}
  (\bibinfo{year}{2003}): \emph{\bibinfo{title}{Algorithms for Ordinal
  Arithmetic}}.
\newblock In \bibinfo{editor}{Franz \surnamestart Baader\surnameend}, editor:
  {\slshape \bibinfo{booktitle}{19th International Conference on Automated
  Deduction ({CADE})}}, {\slshape \bibinfo{series}{Lecture Notes in Computer
  Science}} \bibinfo{volume}{2741}, \bibinfo{publisher}{Springer}, pp.
  \bibinfo{pages}{243--257}, \doi{10.1007/978-3-540-45085-6\_19}.

\bibitemdeclare{inproceedings}{ManoliosVroon04}
\bibitem{ManoliosVroon04}
\bibinfo{author}{Panagiotis \surnamestart Manolios\surnameend} \&
  \bibinfo{author}{Daron \surnamestart Vroon\surnameend}
  (\bibinfo{year}{2004}): \emph{\bibinfo{title}{Integrating Reasoning About
  Ordinal Arithmetic into {ACL2}}}.
\newblock In \bibinfo{editor}{Alan~J. \surnamestart Hu\surnameend} \&
  \bibinfo{editor}{Andrew~K. \surnamestart Martin\surnameend}, editors:
  {\slshape \bibinfo{booktitle}{5th International Conference on Formal Methods
  in Computer-Aided Design ({FMCAD)}}}, {\slshape \bibinfo{series}{Lecture
  Notes in Computer Science}} \bibinfo{volume}{3312},
  \bibinfo{publisher}{Springer}, pp. \bibinfo{pages}{82--97},
  \doi{10.1007/978-3-540-30494-4\_7}.

\bibitemdeclare{article}{MV05}
\bibitem{MV05}
\bibinfo{author}{Panagiotis \surnamestart Manolios\surnameend} \&
  \bibinfo{author}{Daron \surnamestart Vroon\surnameend}
  (\bibinfo{year}{2005}): \emph{\bibinfo{title}{Ordinal Arithmetic: Algorithms
  and Mechanization}}.
\newblock {\slshape \bibinfo{journal}{Journal of Automated Reasoning}}
  \bibinfo{volume}{34}(\bibinfo{number}{4}), pp. \bibinfo{pages}{387--423},
  \doi{10.1007/s10817-005-9023-9}.

\bibitemdeclare{inproceedings}{ccg}
\bibitem{ccg}
\bibinfo{author}{Panagiotis \surnamestart Manolios\surnameend} \&
  \bibinfo{author}{Daron \surnamestart Vroon\surnameend}
  (\bibinfo{year}{2006}): \emph{\bibinfo{title}{Termination Analysis with
  Calling Context Graphs}}.
\newblock In \bibinfo{editor}{Thomas \surnamestart Ball\surnameend} \&
  \bibinfo{editor}{Robert~B. \surnamestart Jones\surnameend}, editors:
  {\slshape \bibinfo{booktitle}{Computer Aided Verification, 18th International
  Conference, {CAV} 2006, Seattle, WA, USA, August 17-20, 2006, Proceedings}},
  {\slshape \bibinfo{series}{Lecture Notes in Computer Science}}
  \bibinfo{volume}{4144}, \bibinfo{publisher}{Springer}, pp.
  \bibinfo{pages}{401--414}, \doi{10.1007/11817963\_36}.

\bibitemdeclare{book}{isabelle-hol}
\bibitem{isabelle-hol}
\bibinfo{author}{Tobias \surnamestart Nipkow\surnameend},
  \bibinfo{author}{Lawrence~C. \surnamestart Paulson\surnameend} \&
  \bibinfo{author}{Markus \surnamestart Wenzel\surnameend}
  (\bibinfo{year}{2002}): \emph{\bibinfo{title}{Isabelle/HOL - {A} Proof
  Assistant for Higher-Order Logic}}.
\newblock {\slshape \bibinfo{series}{Lecture Notes in Computer Science}}
  \bibinfo{volume}{2283}, \bibinfo{publisher}{Springer},
  \doi{10.1007/3-540-45949-9}.

\bibitemdeclare{article}{rswindow}
\bibitem{rswindow}
\bibinfo{author}{Peter~J. \surnamestart Robinson\surnameend} \&
  \bibinfo{author}{John \surnamestart Staples\surnameend}
  (\bibinfo{year}{1993}): \emph{\bibinfo{title}{Formalizing a Hierarchical
  Structure of Practical Mathematical Reasoning}}.
\newblock {\slshape \bibinfo{journal}{Journal of Logic and Computation}}
  \bibinfo{volume}{3}(\bibinfo{number}{1}), pp. \bibinfo{pages}{47--61},
  \doi{10.1093/logcom/3.1.47}.

\bibitemdeclare{inproceedings}{mizar92}
\bibitem{mizar92}
\bibinfo{author}{Piotr \surnamestart Rudnicki\surnameend}
  (\bibinfo{year}{1992}): \emph{\bibinfo{title}{An overview of the Mizar
  project}}.
\newblock In: {\slshape \bibinfo{booktitle}{Proceedings of the 1992 Workshop on
  Types for Proofs and Programs}}.

\bibitemdeclare{misc}{repo}
\bibitem{repo}
\bibinfo{author}{Andrew~T. \surnamestart Walter\surnameend},
  \bibinfo{author}{Ankit \surnamestart Kumar\surnameend} \&
  \bibinfo{author}{Panagiotis \surnamestart Manolios\surnameend}:
  \emph{\bibinfo{title}{Calculational Proof Checker repository}}.
\newblock
  \bibinfo{howpublished}{\url{https://gitlab.com/acl2s/proof-checking/calculational-proof-checker}}.

\bibitemdeclare{misc}{arxiv-hpc-paper}
\bibitem{arxiv-hpc-paper}
\bibinfo{author}{Andrew~T. \surnamestart Walter\surnameend},
  \bibinfo{author}{Ankit \surnamestart Kumar\surnameend} \&
  \bibinfo{author}{Panagiotis \surnamestart Manolios\surnameend}
  (\bibinfo{year}{2023}): \emph{\bibinfo{title}{Calculational Proofs in
  ACL2s}}.
\newblock \eprint{2307.12224}.

\bibitemdeclare{inproceedings}{walter-acl2-systems-programming}
\bibitem{walter-acl2-systems-programming}
\bibinfo{author}{Andrew~T. \surnamestart Walter\surnameend} \&
  \bibinfo{author}{Panagiotis \surnamestart Manolios\surnameend}
  (\bibinfo{year}{2022}): \emph{\bibinfo{title}{{ACL2s} Systems Programming}}.
\newblock In: {\slshape \bibinfo{booktitle}{Proceedings of the Seventeenth
  International Workshop on the {ACL2} Theorem Prover and its Applications}},
  \bibinfo{series}{{EPTCS}}, \doi{10.4204/EPTCS.359.12}.

\bibitemdeclare{article}{isar}
\bibitem{isar}
\bibinfo{author}{Makarius \surnamestart Wenzel\surnameend}
  (\bibinfo{year}{2007}): \emph{\bibinfo{title}{Isabelle/Isar--a generic
  framework for human-readable proof documents}}.
\newblock {\slshape \bibinfo{journal}{Studies in Logic, Grammar and Rhetoric}}
  \bibinfo{volume}{10}(\bibinfo{number}{23}).
\newblock \urlprefix\url{http://mizar.org/trybulec65}.

\bibitemdeclare{misc}{xtext}
\bibitem{xtext}
\bibinfo{author}{\surnamestart {Xtext Contributors}\surnameend}:
  \emph{\bibinfo{title}{Xtext}}.
\newblock \urlprefix\url{https://www.eclipse.org/Xtext/}.
\newblock \bibinfo{note}{Accessed on April 25th, 2022}.

\end{thebibliography}
\end{document}